\documentstyle[preprint,aps,epsf]{revtex}

\newcommand{\be}{\begin{equation}}
\newcommand{\ee}{\end{equation}}

\begin{document}

\title{ Aharonov-Bohm and Aharonov-Casher Effects: 
          Connections to Dynamics of Topological Singularities }

\author{ P. Ao$^1$ and Q. Niu$^2$ }

\address{ $^1$Departments of Theoretical Physics, 
          Ume\aa{\ }University, S-901 87, Ume\aa, Sweden \\
          $^2$Department of Physics, University of Texas, Austin, TX 78712 }


\maketitle

We analyze the physical processes involved 
in the Aharonov-Bohm (A-B) and the Aharonov-Casher (A-C) effects, 
showing that an incomplete A-B effect knowledge can lead
a totally wrong conclusion on the A-C effect.
Based on this we demonstrate that the Magnus force, the net force, 
is the only transverse force on a moving vortex, 
in analogous to the net charge in A-C effect. 
This conclusion has been arrived both theoretically and experimentally.

Let us begin with a well accepted situation, 
the connection between the A-B effect\cite{ab}
 and the A-C effect\cite{ac}.
The A-C phenomenon is the effect of charges on a moving 
magnetic flux line, 
the dual effect of A-B phenomenon of the magnetic flux line effect on
moving charges.
Within the non-relativistic formulation, 
those two effects can be formally
related to each other by a Galilean transformation. 
However, the crucial difference between A-B and A-C effects in
real calculations is that, for the A-B effect a moving charge `sees' the
net magnetic flux, but for the A-C effect a moving magnetic flux line
`sees' the net charge.
It is rather easy to conceive the situation of a flux line 
in a condensed matter: suppose one knows
perfectly well a large  A-B effect for conduction electrons, can he draw 
a definite conclusion on the A-C effect for the moving  flux line?
The answer is NO,
because the net charge for the A-C effect comes from various contributions,
and the contribution from conduction electrons is only one of them. 
For example, the A-C effect can be zero in the case of charge neutrality, 
and, be either 
positive or negative depending on the net charge feel by the flux line.
Until all charges, valence electrons, ionic background, quarks, ....,
have been considered, the A-B effect for conduction electrons alone
can not yield any information on the A-C effect for a moving flux line.
This illustrates the pitfall in using the incomplete A-B effect 
information to deduce the A-C effect.
The best way to find the effect of net charges on 
a moving magnetic flux line is a direct calculation following
this flux line, as already suggested by Aharonov and Casher.

Now we are ready to discuss the transverse force on a moving vortex.
The phonons act like, which may sounds natural to some authors, 
conduction electrons, and a vortex like a magnetic flux line. 
Let's accept this analogy without further question, 
though it may not be true.
 The phonons scattering off a vortex
resembles the A-B effect, while a moving vortex picks a possible transverse
effect from phonons resembles the A-C effect.
According to above analysis of the connection between A-B and A-C effects,
the transverse effect due to phonons cannot tell us 
anything about the net transverse force felt by a moving vortex.
One has to consider the normal fluid, superfluid, the total fluid, and other
contributions involved.
This phonon effect can be much larger than the net transverse force
on a vortex, and can even carry a negative sign.
In analogous to the calculation of A-C effect,
the exact result, obtained by calculation the Berry 
phase\cite{at}, or, explicit counting individual state 
contributions\cite{tan}, shows the magnitude of 
the net transverse force on a moving vortex 
is determined by the superfluid 
density. The absence of any additional transverse force 
has also been shown in a recent path integral derivation
of vortex dynamics.\cite{za}

We should emphasize that 
there are many possible individual contributions to
the transverse force. 
They do not all carry the same sign.
The topology of the vortex requires that their sum, the net effect,
is universal, determined only by the superfluid, although
any one of such contributions may have a magnitude larger than the net 
effect.
Simultaneously, there are 
equal number or more contributions to the longitudinal force, the 
friction. The important fact is that they all carry
the same sign. 
Hence one has to sum all of them up to determine the friction.
This is a detail-sensitive procedure which shows friction has no
 universal value.
This striking difference
between the transverse force and friction may have caused 
a considerable amount of confusion in the literature
to understand the exact result on the transverse force. 

It is also worthwhile to emphasize that, in addition to
above theoretical analysis, well accepted experiments,
such as 
the vibrating wire\cite{vinen}, the vortex procession\cite{zieve}, 
as well as a recent vibrating reed experiment done in 
Ume\aa\cite{zbs},
have shown the non-existence of extra transverse forces.

Finally, we point out that the transverse force in the A-B effect has 
been well established\cite{hedegard}, and routinely used in the study of 
quantum Hall effect\cite{niu} and vortex dynamics in Josephson junction 
arrays\cite{zta}.

The conclusion is, 
the analogy to the Aharonov-Bohm and Aharonov-Casher 
effects has put the recent topological demonstration of the vortex dynamics 
on a firmer ground. 
There is no extra transverse force 
other than the Magnus force, shown both theoretically and experimentally.

\end{document}